\documentclass[prl,amsmath,amssymb,twocolumn,superscriptaddress]{revtex4}
\usepackage{amsmath, amstext, amssymb, amsfonts, amsxtra}
\usepackage[usenames]{color}
\usepackage{grffile}
\usepackage{textcomp}
\usepackage{xspace}
\usepackage{bbm}
\usepackage{bm}
\newcommand{\be}{\begin{equation}}
\newcommand{\ee}{\end{equation}}

\newcommand{\xiamen}{Department of Physics and Key Laboratory of Low Dimensional
Condensed Matter Physics (Department of Education of Fujian Province), Xiamen
University, Xiamen 361005, Fujian, China}
\newcommand{\lanzhou}{Lanzhou Center for Theoretical Physics, Lanzhou University,
Lanzhou 730000, Gansu, China}
\newcommand{\como}{Center for Nonlinear and Complex Systems, Dipartimento di
Scienza e Alta Tecnologia, Universit\`a degli Studi dell'Insubria,
via Valleggio 11, 22100 Como, Italy}
\newcommand{\infn}{Istituto Nazionale di Fisica Nucleare, Sezione di Milano,
via Celoria 16, 20133 Milano, Italy}
\newcommand{\brazil}{International Institute of Physics, Federal University
of Rio Grande do Norte, Campus Universit\'ario - Lagoa Nova, CP. 1613, Natal,
Rio Grande Do Norte 59078-970, Brazil}
\newcommand{\NEST}{NEST, Istituto Nanoscienze-CNR, I-56126 Pisa, Italy}

\begin{document}

\title{Classical Physics and Blackbody Radiation}

\author{Jiao Wang}
\affiliation{\xiamen}
\affiliation{\lanzhou}
\author{Giulio Casati}
\affiliation{\como}
\affiliation{\brazil}
\author{Giuliano Benenti}
\affiliation{\como}
\affiliation{\infn}
\affiliation{\NEST}

\begin{abstract}
We investigate the properties of the blackbody spectrum by direct numerical solution of
the classical equations of motion of a one-dimensional model that contains the essential
general features of the field-matter interaction. Our results, which do not rely on any
statistical assumption,  show that the classical blackbody spectrum exhibits remarkable
properties: (i) a quasistationary state characterized by scaling properties, (ii) consistency
with the Stefan-Boltzmann law, and (iii) a high-frequency cutoff. Our work is a preliminary 
step in the understanding of statistical properties of infinite dimensional systems.
\end{abstract}

\maketitle

\emph{Introduction.--}
The inability of classical physics to account for the experimentally observed frequency
spectrum of blackbody radiation is at the origin of quantum theory. In spite of desperate 
attempts, the falloff of the
blackbody curve at high frequencies could not be explained by classical mechanics. To be
more precise, however, this is not a failure of classical theory \emph{per se}. Indeed,
it is the classical theory with the additional assumption of energy equipartition which
leads to the Rayleigh-Jeans  radiation  law,  implying the unphysical ultraviolet
catastrophe~\cite{Jeans16}.

On the other hand, classical ergodic theory, from which equipartition theorem follows,
is valid only for systems with a {\it finite} number of degree of freedoms. Ergodicity,
that is, independence of time averages on initial conditions, does not imply equipartition
of energy for a system with infinite degrees of freedom, since there is no invariant measure
at hand to define a microcanonical ensemble. A main difficulty here stems from the fact that
the radiation field has $N_m=\infty$ degrees of freedom (modes) and that the two limits
$N_m \rightarrow \infty$ and time $ t \rightarrow\infty$ do not commute.
Therefore, the question of what classical mechanics predicts on the properties of the
radiation field in equilibrium with matter is still open.

Even more interesting is the Stefan-Boltzmann law which states that at temperature $T$ the
total radiation energy $ E \propto T^4$. It is remarkable that this formula, which is well
in agreement with experimental data, was derived by Boltzmann in 1884 on
\textit{purely classical thermodynamics basis}~\cite{Boltzmann95}. As such, it should be
a consequence of the classical equations of motion. Notice that this result is formally
in contradiction with the Rayleigh-Jeans law, according to which $E\propto T $.

What about the solution of  the nonlinear classical Newton-Maxwell equations of motion that
govern the field-matter interaction? Will dynamics agree and to what extent, with the above
classical statistical and thermodynamics predictions? After 150 years, this
question concerning a fundamental problem in the development of modern physics remains unsolved.

Here, we numerically integrate the exact classical equations of motion of a blackbody model,
without any statistical assumption, and we show that classical dynamics leads to a quasistationary
state that is consistent with the Stefan-Boltzmann law
and  where  the energy  distribution  over  the  field  normal  modes exhibits  an exponentially
decreasing tail. Therefore the solution of the classical equations of motion naturally leads
to a high-frequency  cutoff  of  the  blackbody  spectrum.

\emph{Model.-}
To investigate the dynamics of charged particles interacting with the electromagnetic field
in a cavity is a formidable task. Here we consider a variant of a model introduced long
ago~\cite{Bocchieri72} and afterwards studied in several papers~\cite{Bocchieri74, Giulio77,
Galgani82,Giulio83, Livi87, Alabiso89, EPL99}. In spite of its simplicity,  our model retains
the general essential features of field-matter interaction: (i) the field modes can only exchange
energy via interaction with matter and (ii) the free electromagnetic field in the cavity is by
itself a linear system and nonlinearity is provided by matter.

\begin{figure}[t]
\includegraphics[width=9.5cm]{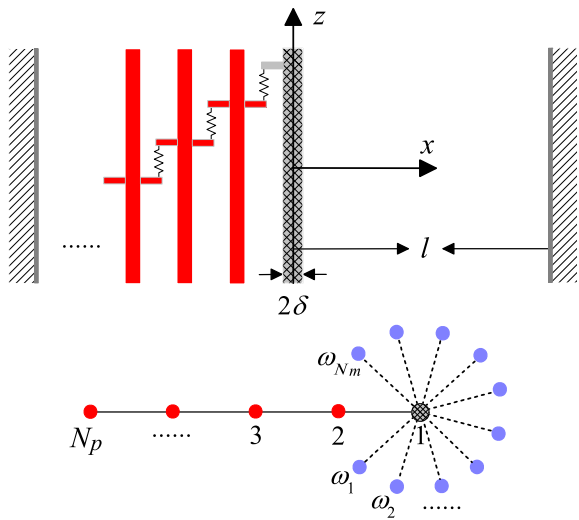}
\caption{Schematic drawing (top) of the classical radiant cavity model. A charged plate
of width $2\delta$ is placed at the center of the cavity bounded by two mirrors a distance
$2l$ apart. The charged plate interacts with a fixed number of neutral plates (to the left
in the figure) and with the normal modes of the electromagnetic field. The topology of
interactions is also shown (bottom), where the gray dot, red dots, and blue dots are for
the charged plate, the neutral plates, and the field modes, respectively.}
\label{fig:scheme}
\end{figure}

A sketch of our model is drawn in Fig.~\ref{fig:scheme}. It consists of an electromagnetic
field confined in between two parallel, perfectly reflecting plane mirrors, a distance $2l$
apart. We take Cartesian coordinates $xyz$ with the $x$ axis normal to the mirrors and
restrict to excitations only dependent on $x$, thus getting a one-dimensional radiant cavity,
the normal modes of which have angular frequencies $\omega_k$.  We then introduce $N_p$
coupled plates, which are all parallel to the mirrors and which are constrained to move
only in the $z$ direction. Only the plate positioned midway between the mirrors is charged
and therefore interacts with the normal modes of the field.
We denote by $\sigma$ and $m$ the charge and mass densities per unit surface of the plate,
and $f(x)$ is the normalized (transverse) distribution of charge in the plate, whose thickness
is $2\delta$.

The Hamiltonian of the full system, plates plus field, can be written as
\begin{eqnarray}
H & = & \sum_{j=2}^{N_p} \left[\frac{P^2_j}{2m}+\tilde V(z_j)+V(z_{j-1},z_j)\right]
+\tilde V(z_1)\nonumber\\
&+&\frac{1}{2m}(P_1-\varepsilon\sum_{k=1}^{\infty} a_k q_k)^2
+\frac{1}{2}\sum_{k=1}^{\infty} (p_k^2+\omega_k^2 q_k^2),
\end{eqnarray}
where $P_j$ and $z_j$ are the conjugate momentum and displacement of the $j$th plate,
$\tilde V$ and $V$ are the on site and interacting potentials of plates, and $p_k$ and
$q_k$ are conjugate variables of the $k$th normal mode of the field with frequency $\omega_k$.
Note that the field normal modes are infinite but, as we shall see below, only a finite, small
number $N_m(t)$ of them are actually involved during the computation time. In Eq.~(1), the 
term $\frac{1}{2m}(P_1-\varepsilon\sum_k a_k q_k)^2$
accounts for the interaction between the charged plate and the normal modes, where
$\varepsilon=2\sigma\sqrt{\pi/l}$
is the matter-field coupling parameter. We assume that $f(x)$ is an even function, and therefore,
even normal modes do not interact with the charged plate, while for an odd normal mode with wave
number $2k-1$, $a_k=\int_{-\delta}^{\delta} f(x) \cos \frac{\omega_k x}{c}dx$, where $c$ is the
speed of light and $\omega_k =\frac{\pi c}{2l}(2k-1)$.

If the charge is removed, $\varepsilon=0$, the plates and the modes are decoupled.
We choose, as interaction potential among the matter degrees of freedom,
the lattice $\phi^4$ model~\cite{Aoki-phi4,Parisi-phi4}
\begin{eqnarray}
\tilde V(z_j)=\frac{1}{4}\,\gamma z_j^4, ~~
V(z_{j-1},z_j)=\frac{1}{2} \,\kappa (z_j-z_{j-1})^2.
\end{eqnarray}
We checked that, for $N_p>4$ and for an average energy per plate larger than $0.1$
(in our units $l=\pi$, $c=m=\textcolor{blue}{\sigma=1}$, and the Boltzmann constant $k_B=1$; moreover
we set $\gamma=\kappa=1$), this model is chaotic with $N_p$ positive Lyapunov exponents.

Initially, all the energy is assigned to matter only (the plates), and we focus on how
the energy is transferred and distributed among the field normal modes. For the charge
distribution we choose
$f(x)=A \exp(\frac{\delta^2}{x^2-\delta^2})$ ($|x|<\delta$) (a
compactly supported $C^\infty$ function, with $A$ normalization constant,
$A=45.04...$ for the width $\delta=0.05$ used in our simulations).
We have verified that, qualitatively, the results
do not depend on the particular choice of the charge distribution, provided it is a smooth
function~(see Supplemental Material~\cite{supplemental}).
In addition, we take $N_p=16$ and we have verified that the whole system remains chaotic during
the entire relaxation process accessible to simulations ($t\sim 10^9$)~\cite{supplemental}.

\begin{figure}[!t]
\includegraphics[width=8.5cm]{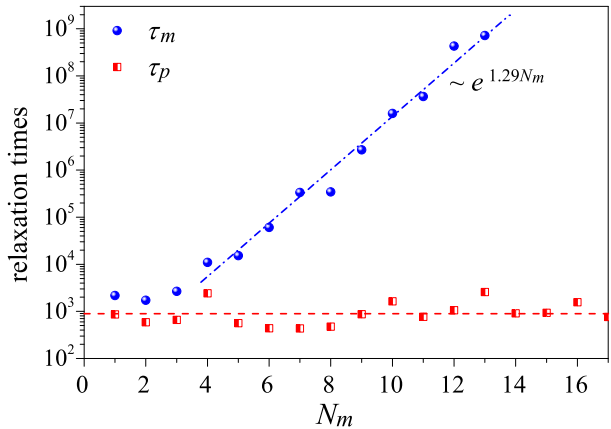}
\caption{The relaxation time of plates ($\tau_p$) and of both plates and field modes ($\tau_m$),
where $N_m$ field modes are considered in the simulation. Here the number of plates $N_p=16$,
and the total energy $E_{tot}=2^4$ is initially assigned to the charged plate as kinetic energy.
For other initial conditions where the total energy is distributed randomly among the plates,
$\tau_m\sim e^{1.29N_m}$ still holds.}
\label{fig:relaxation}
\end{figure}

\emph{Results.--}
Since our simulations are
for energies well above the stochasticity threshold, for any \emph{finite}
number of field modes, equipartition is expected among the degrees of freedom. Our first step
is therefore to investigate how long it takes for equipartition to set in.
As shown in Fig.~\ref{fig:relaxation},
equipartition is reached among the plates after a relatively short time, $\tau_p$, that,
basically, does not depend on the number $N_m$ of field modes. Instead, global equipartition
among field and matter degrees of freedom is reached after a time $\tau_m$ that exponentially
increases with $N_m$. The best fitting suggests  $\tau_m\sim e^{1.29N_m}$, in clear contrast with
$\tau_p \approx 10^3$~\cite{supplemental}.
If we extrapolate this exponential dependence of the equipartition time $\tau_m$ to larger $N_m$
values, then it turns out that already for $N_m\approx 48$ field normal modes, a time of the order
of the age of the Universe is required in order to reach equipartition (in a cavity of $l\sim 1$
m). Note that the two limits $t\rightarrow \infty$ and $N_m\rightarrow \infty$ do not commute. If one
takes the limit $t\rightarrow \infty$ first, then one will always get equipartition while for the
blackbody problem it is necessary to take the limit $N_m\rightarrow \infty$ first.  In order to
simulate this latter situation, in our computations we take $N_m =120$, which, in view of the
results of Fig.~\ref{fig:relaxation}, is sufficiently large to neglect the higher normal modes.

\begin{figure}[!]
\includegraphics[width=8.5cm]{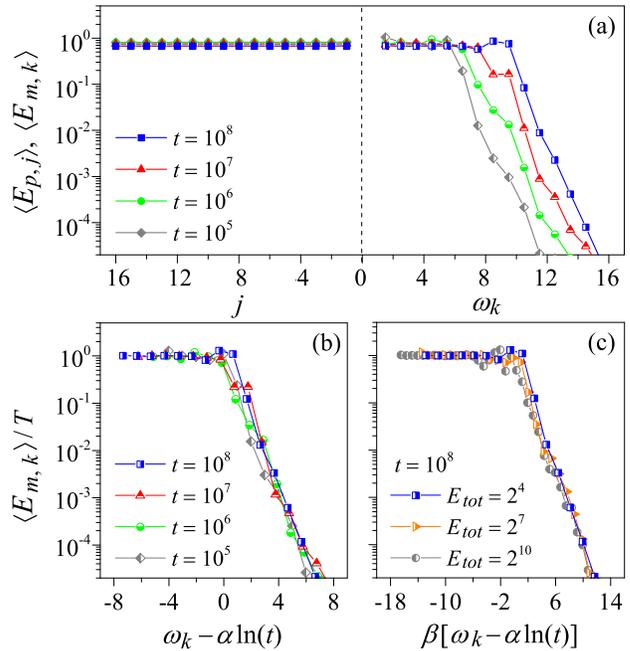}
\caption{(a) The average energy of the plates (full symbols) and of the field modes (half filled
symbols), for total energy $E_{tot}=2^4$ at different times. (b) $\langle E_{m,k}\rangle$ scaled
by the temperature of plates and the mode frequencies being shifted by $\alpha$ln$(t)$. (c)
$\langle E_{m,k} \rangle$ versus the scaled and shifted frequencies at a given time. The results
for different total energies collapse on the same curve. The energy of the system is initially
randomly assigned to the plates as kinetic energy.}
\label{fig:scaling}
\end{figure}

We now turn to the energy distribution over the field normal modes. In Fig.~\ref{fig:scaling}(a)
we plot the average energy $\langle E_{m,k}(t) \rangle$ of the normal modes over their frequency.
The main feature of the distribution $\langle E_{m,k}\rangle$ is the presence of a plateau
followed by a rapid, exponential decay.
The average value $\langle E_{m,k}\rangle$ of the plateau normal mode energies is equal to the
average energy $\langle E_{p,j}\rangle$ of the plates. This suggests that the field modes in
the plateau are thermalized with matter, at temperature $T=\langle E_{p,j}\rangle$. This
thermalization process evolves in time very slowly; indeed, as seen in Fig.~\ref{fig:scaling}(a),
it takes about 3 orders of magnitude in time (from $10^5$ to $10^8$) in order to increase
only by four the number of thermalized modes. This observation is consistent with the results of
Fig.~\ref{fig:relaxation} and suggests that the system relaxes to a quasistationary state after
which it evolves very slowly, logarithmically in time.


It is remarkable that the quasistationary state has a clear scaling property. This is shown in
Figs.~\ref{fig:scaling}(b) and \ref{fig:scaling}(c), where we plot $\langle E_{m,k}(t)\rangle/T(t)$
over $[\omega_k - \alpha\ln(t)]$ and over $\beta[\omega_k - \alpha\ln(t)]$, where $\alpha$ and
$\beta$ depend on the total energy only, $\alpha=0.23E_{tot}^{0.24}$ and $\beta=3.22E_{tot}^{-0.22}$
based on the best fitting of the data for $10^4 \le t \le 10^8$ and $2^4 \le E_{tot} \le 2^{10}$.

\begin{figure}[!]
\includegraphics[width=8.5cm]{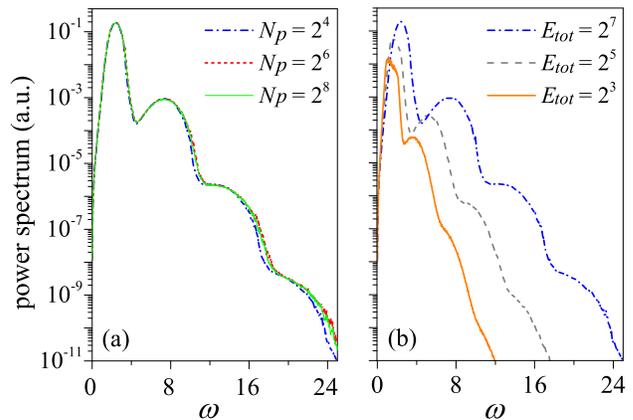}
\caption{Power spectrum of the time series $P_1(t)$ for the momentum of the
charged plate, in a system of $N_p$ plates with total energy $E_{tot}$, for (a)  $E_{tot}/N_p=8$
and (b) $N_p=16$.}
\label{fig:powerspectrum}
\end{figure}

We cannot give a rigorous explanation for the appearance of an exponential
cutoff. Nevertheless, we can provide an intuitive understanding by observing
(see Fig.~\ref{fig:powerspectrum}) that, as is typically the case for nonlinear interacting systems,
 the power spectrum of the charged plate has a finite frequency band, followed by an exponential tail. 
 For the field modes within this band, thermalization
with matter is achieved rapidly. On the other hand, for the field modes with frequencies in the
exponential tail, thermalization requires exponentially long timescales. We can see from
Fig.~\ref{fig:powerspectrum} that the power spectrum is, in practice, independent of the number
$N_p$ of plates (for a fixed energy per plate), while for a given $N_p$ the bandwidth increases
with the total energy $E_{tot}$. Such dependence is consistent with the reduction of the thermalization
times (up to a given frequency) with increasing $E_{tot}$.

\begin{figure}[!]
\includegraphics[width=8.5cm]{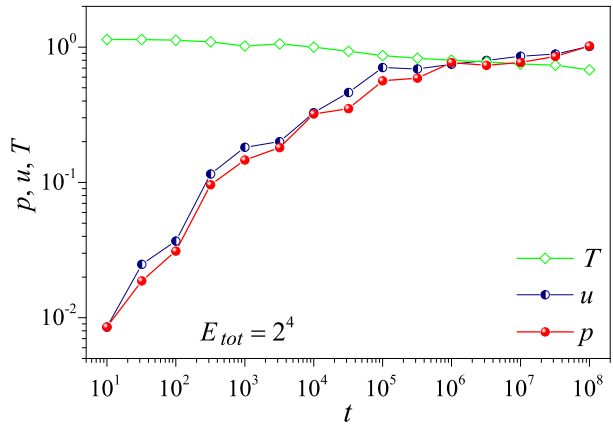}
\caption{The pressure $p$, the energy density $u$ of the cavity, and the temperature $T$ of
the plates as functions of time.}
\label{fig:pressure}
\vskip-0.15cm
\end{figure}

\begin{figure}[!]
\includegraphics[width=8.5cm]{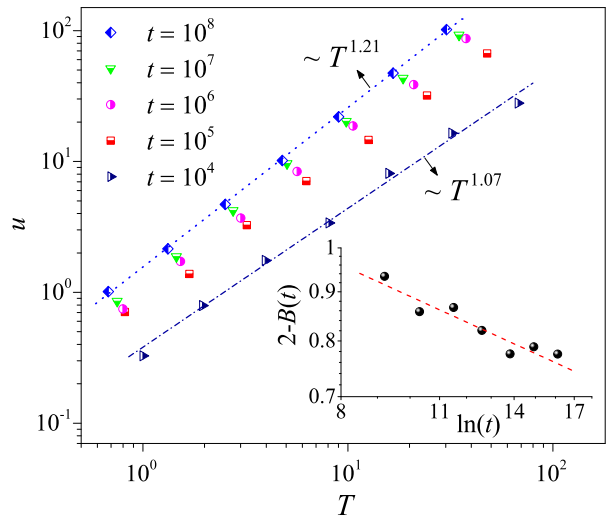}
\caption{The energy density $u$ in the cavity versus the temperature $T$ of the plates at
various evolution times. At a given time, the seven data points correspond to total energies
$E_{tot}=2^4, 2^5, ..., 2^{10}$, respectively. At each time the data exhibit a power law relation
$u\propto T^B$, and the best fitting  shows that the power law exponent $B(t)$ slowly changes from
1.07 to 1.21 as time changes by 4 orders of magnitude. The fitting line in the inset gives the
expression $2-B(t)\sim(\ln t)^{-0.33}$.}
\vskip-0.15cm
\label{fig:SB}
\end{figure}

Note that the total energy of the modes in the exponential tail is negligible compared with
that of the thermalized modes on the plateau. Therefore, it is tempting to conjecture that the
quasistationary state should be approximately described by equilibrium thermodynamics. To this
end, following Boltzmann, we start from the fundamental thermodynamic relation $dU=TdS-pdV$,
which implies
\begin{eqnarray}
\frac{\partial U}{\partial V}{\Bigg |}_T=-p+T\frac{\partial p}{\partial T}{\Bigg |}_V,
\end{eqnarray}
where $U= U(T,V)$ is the total energy, $S$ is the entropy, $p$ is the pressure of the radiation
field, $V$ is the volume of the cavity, and $T$ is the temperature of the plates. The relation between
the energy of the electromagnetic radiation and its pressure has been derived by Boltzmann and
for the one-dimensional case reads:
\begin{eqnarray}
p= u,
\end{eqnarray}
where $u=U/V$ is a universal function of temperature and independent of volume. From Eq.~(3)
 one then has
\begin{eqnarray}
u(T)=CT^{2},
\end{eqnarray}
where $C$ is a constant. This is the Stefan-Boltzmann law in one dimension. Quite remarkably,
these predictions based on equilibrium thermodynamics are consistent with our model in the
quasistationary state. In Fig.~\ref{fig:pressure}, we plot $p$, $u$, and $T$ versus the evolution
time for $E_{tot}=2^4$. Here $p$ is numerically computed~\cite{supplemental}
and $T =\langle E_{p,j}\rangle$ is the temperature of matter. It can be noticed that the
relation (4) between energy density and pressure sets in almost immediately.
In Fig.~\ref{fig:SB}, we plot $u$ versus $T$ for various values of the total energy of the
system. It is seen that $u\propto T^{B}$ where the exponent $B(t)$ increases from 1.07 to 1.21,
as time $t$ goes from $10^4$ to $10^8$. This dependence is compatible with Eq.~(5). Indeed the
inset of Fig.~\ref{fig:SB} suggests that the value 2 might be approached logarithmically in time.

\emph{Summary and discussion.--}
In summary, we have studied a model of a classical radiant cavity.
If one considers a fixed, finite number $N_m$ of field modes, then the system, as expected,
approaches equipartition, even though the relaxation times increase exponentially with the
number $N_m$ of modes. If instead one takes the limit $N_m \rightarrow \infty$ first, as it
is required in order to study the blackbody radiation, then the system relaxes to a
quasistationary state characterized by a low frequency plateau, followed  by a high
frequency exponential cutoff. The computed field energy density turns out to be consistent
with the Stefan-Boltzmann law.

Our model contains the basic ingredients of an electromagnetic field interacting with matter.
Since interaction among the field modes is mediated by mechanical degrees of
freedom, in general, we expect thermalization of the field modes with matter to be effective
only within the frequency bandwidth of mechanical motion, with slow thermalization outside such
bandwidth. We therefore conjecture that the appearance of a time-dependent cutoff is a general
feature of classical dynamics of matter-field interaction. Nevertheless, it would be interesting
to investigate higher-dimensional models or, more generally, other classical field theories. Such
studies could help our understanding of the statistical properties of nonlinear dynamical systems
with infinite degrees of freedom. In this frame, the results presented here on a classical model
can be considered a preliminary step before addressing ergodicity in quantum field theories. This
problem will require a nontrivial extension of concepts and tools recently developed for the
investigation of thermalization and localization in many-body quantum systems~\cite{Polkovnikov11,
Huse15, Borgonovi16, Rigol16, Altman19}.

We acknowledge support by the NSFC (Grants No. 12075198 and No. 12047501)
and by the INFN through the project QUANTUM.

\section{Supplemental Material}

\subsection{Results for other charge distributions}

In this section, we show that the exponentially slow relaxation of the field modes
is a general feature, independent of the particular charge distribution $f(x)$.
For this purpose we focus, in between the other charge distributions we have considered
(apart from the one in the main text),
on two limiting cases: the Dirac delta function and the hump function.
The Dirac-delta function $f(x)=\delta(x)$ is an extreme case, for which
the coefficients $a_k$ are constant. On the other hand, for smooth functions
like the hump function $f(x)=B(\delta^2-x^2)/(\delta^4+x^4)$ or the function
$f(x)=A\exp({\delta^2}/({x^2-\delta^2}))$ considered
in the main text, (for both functions, with a numerical normalization factor in front)
the coefficients $a_k$ decay with $k$ faster than algebraically
(see Fig.~7).

Below we report the results for the relaxation time and the average energy of the plates
and of the field modes, corresponding to Figs.~2 and 3(a) of the main text, but for
the delta and the hump function. While for the hump charge distribution the obtained
results are quite similar to those reported in the main text (see Fig.~8), for the delta-function
distribution the energy transfer to the field modes is more efficient (see Fig.~9). On the
other hand, it appears clearly
that also in this limiting case the average energy of the normal modes drops exponentially
after the thermalization plateau and before the algebraic tail $\propto 1/k^2$ determined by
the coefficients $a_k$ (see below). The exponential decay goes down to smaller and smaller
energy as time increases and more normal modes are thermalized. Moreover, thermalization
of the field modes is again logarithmically slow in time, the relaxation time for $N_m$
normal modes being $\tau_m\sim e^{\alpha N_m}$, even though the coefficient $\alpha=0.54$
is smaller than the value $\alpha=1.29$ obtained for the smooth functions (the hump
distribution and the distribution considered in the main text).

\begin{figure}
\includegraphics[width=8.0cm]{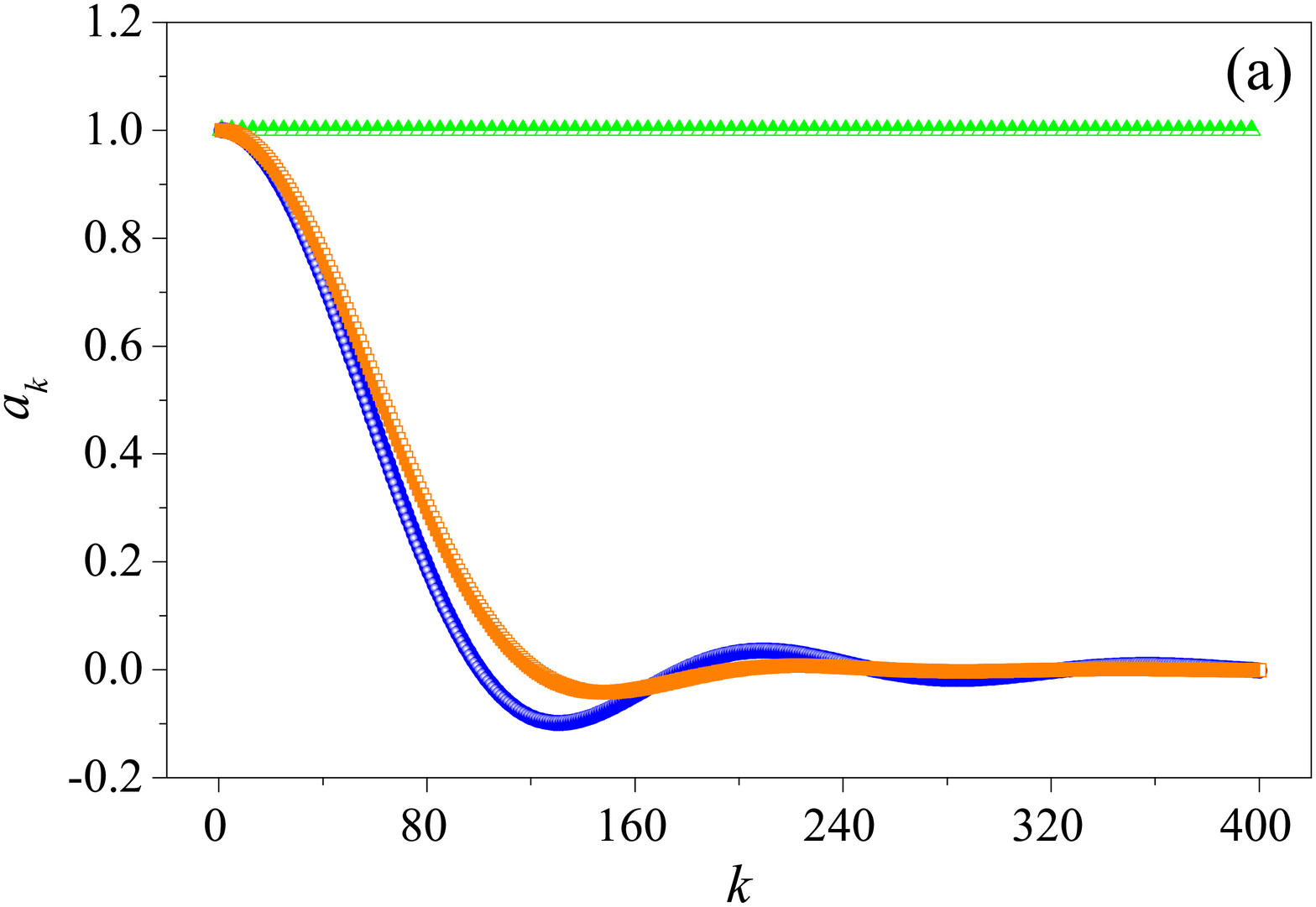}
\includegraphics[width=8.0cm]{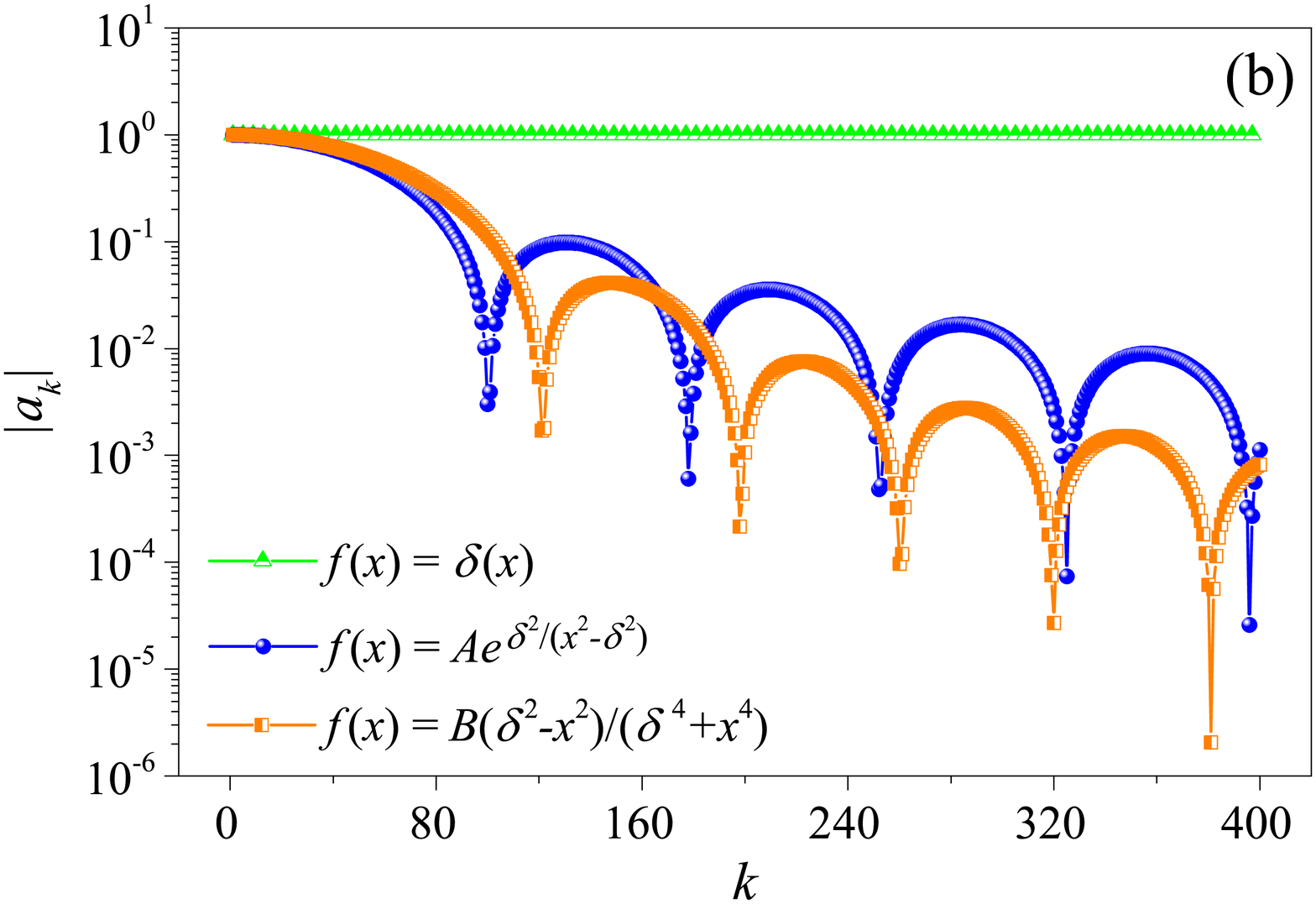}
\caption{Coupling coefficients
$a_k$ (top panel) between the charged plate and the $k$-th field mode,
and their absolute values $|a_k|$ (bottom panel), for
three charge distribution functions considered in our study. The coefficients corresponding to
$f(x)=A\exp(\frac{\delta^2}{x^2-\delta^2})$ and $f(x)=B\frac{\delta^2-x^2}{\delta^4-x^4}$
(in both cases, $\delta=0.05$, and correspondingly $A=???$ and $B=???$)
oscillate around zero and meanwhile decay, while for the Dirac delta function
$a_k=1$, independently of $k$.}
\end{figure}

\begin{figure}
\includegraphics[width=8.0cm]{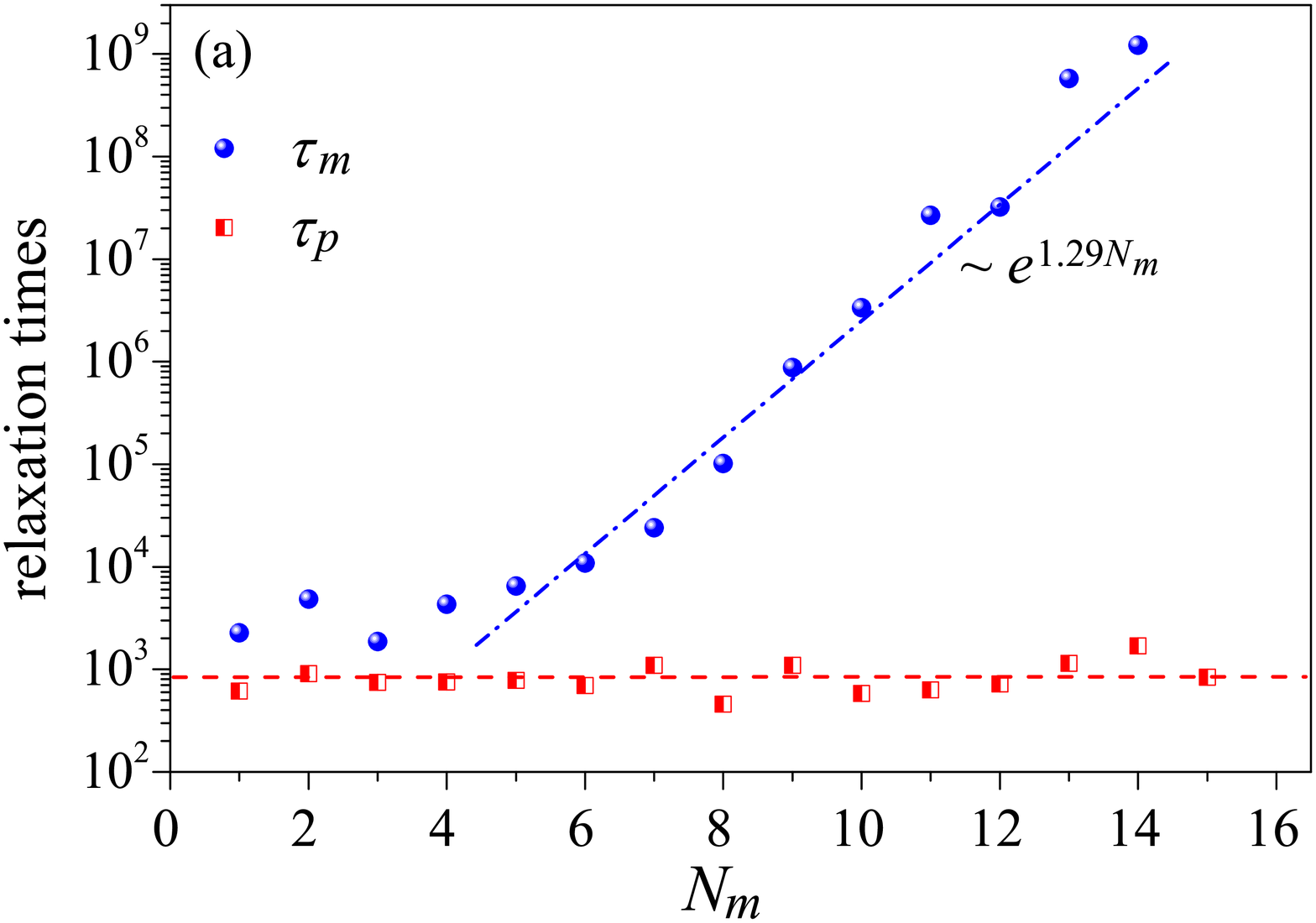}
\includegraphics[width=8.0cm]{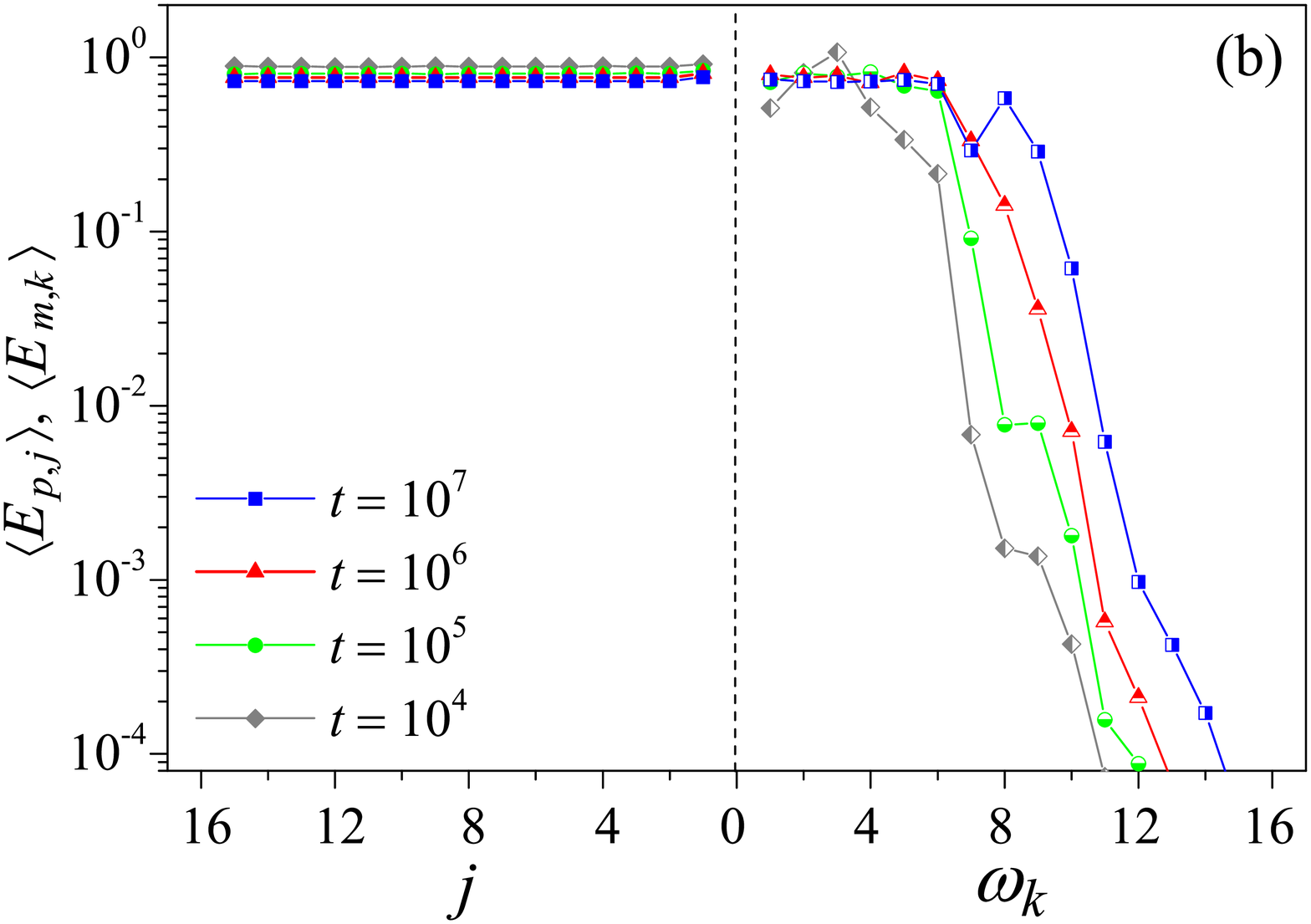}
\caption{The relaxation process for the case where the charge distribution
function is a smooth function, $f(x)=B(\delta^2-x^2)/(\delta^2+x^4)$,
with $\delta=0.05$ and $B=19.51...$. (a) Relaxation time of plates ($\tau_p$) and of both
plates and field modes ($\tau_m$). (b) Average energy of the plates (full
symbols) and of the field modes (half-filled symbols) at different times,
for total energy
$E_{tot}=16$.}
\end{figure}

\begin{figure}
\includegraphics[width=8.0cm]{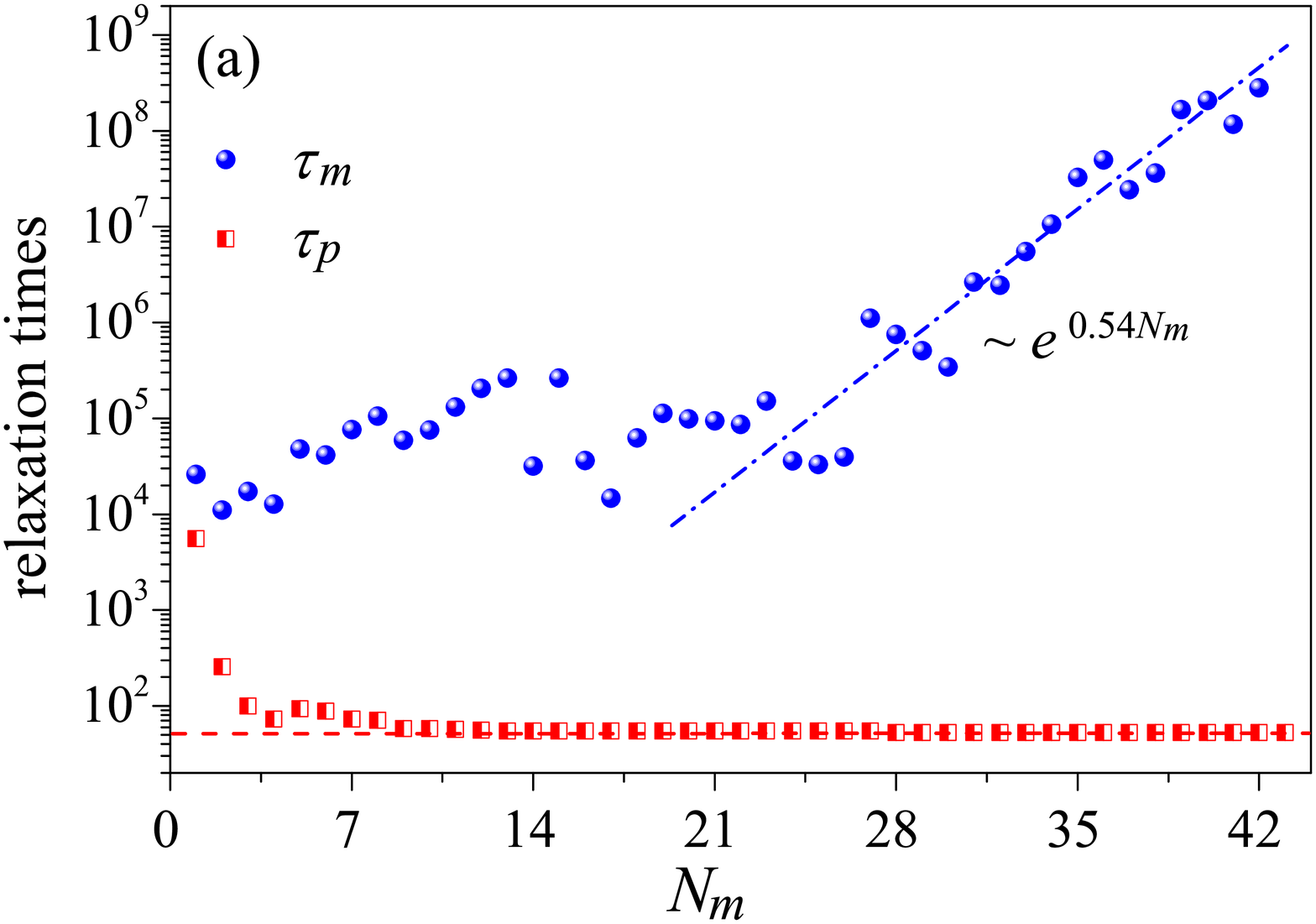}
\includegraphics[width=8.0cm]{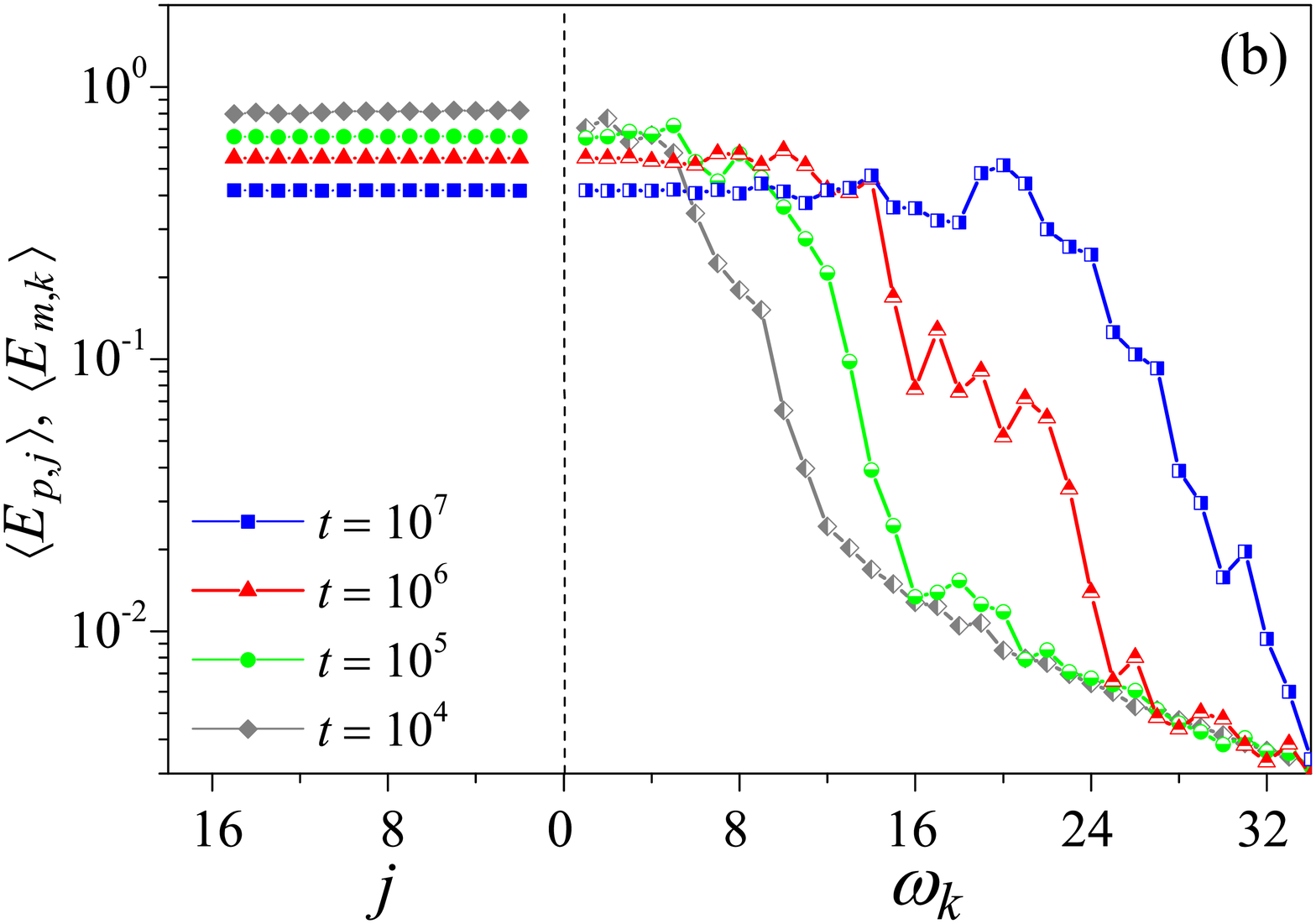}
\caption{As in Fig.~8, but for the limiting case where the charge
distribution function is a Dirac $\delta$-function, $f(x)=\delta(x)$.}
\end{figure}

Finally, we add the explanation of the power-law tail observed in Fig.~9(b).
For that purpose, it is enough to consider the linear model with the charged plate coupled
to the cavity modes. That is, the second line of Eq. (1) of the main text:
\begin{eqnarray}
H_{\rm lin}=\frac{1}{2m}(P_1-\varepsilon\sum_{k=1}^{\infty} a_k q_k)^2
+\frac{1}{2}\sum_{k=1}^{\infty} (p_k^2+\omega_k^2 q_k^2).
\end{eqnarray}
In this case, the state of the system at time $t$, written as an $\infty$-dimensional vector,
\begin{equation}
{\bf q}(t)=\left\{m^{1/2}\dot{z}_1(t),
...,\omega_{k}q_{k}(t),
p_{k}(t),...\right\},
\end{equation}
(the total energy is the square norm of such vector)
can be expanded over the eigenstates of the overall (plate plus cavity) linear system,
\begin{equation}
{\bf u}_{k}=C_k\left\{1,...,
\frac{\varepsilon\,a_{k}\,\omega_{k}}
{\Omega_k^2+\omega_{k}^2}\,,
\frac{\varepsilon\,a_{k}\,\Omega_k}
{\Omega_k^2+\omega_{k}^2},...\right\},
\end{equation}
where $\varepsilon=2\sigma(\pi/ml)^{1/2}$ and $C_k$ is a normalization constant.
These eigenvectors define normal modes of the total system, with eigenfrequencies
$\Omega_k$ given by the imaginary roots of the secular equation:
\begin{equation}
\Omega_k\left(1+\varepsilon^2\sum_{n=1}^{\infty}\,\frac{a_{n}^2}
{\Omega_k^2+\omega_{n}^2}\right)=0.
\end{equation}
In the linear system, there is no thermalization and the field modes are populated as determined
by the Fourier coefficients $a_k$. In particular, for large $k$, the energy on mode $k$ drops as
$1/k^2$ when the charge distribution $f(x)$ is a Dirac delta-function ($a_k=1$ independently of
$k$ in this case), and faster than algebraically
for the hump distribution and the distribution considered in the main text.

On the other hand, such difference does not affect qualitatively the conclusions of our study when
nonlinear terms, mixing the normal modes, are added, inducing exponentially slow thermalization of
the field modes, independently of the particular choice of $f(x)$.


\subsection{Details of numerical simulations}

In our numerics, we have used a 4th order Runge-Kutta algorithm, with
integration time step $\Delta t=10^{-4}$, which we have verified to be
short enough to ensure reliable results up to $t\sim 10^9$ and for total
energy $E_{tot}$ ranging between 16 and 1024.

We have defined the energies $E_{m,k}=p_k^2$ and $E_{p,j}=P_j^2$. In order
to suppress fluctuations,  we perform, for a given initial condition, a
partial time average of $E_{m,k}(t)$ over the time interval $(t,2t)$. In
addition we make an ensemble average (denoted as $\langle ... \rangle$)
over nine initial conditions. For the normal modes of the field, we have
checked that $\langle p_k^2\rangle=\omega_k^2 \langle q_k^2\rangle$. For
the neutral plates, using the equipartition theorem we obtain
\begin{equation}
\left\langle P_j\frac{\partial H}{\partial P_j}\right\rangle=
\left\langle z_j\frac{\partial H}{\partial z_j}\right\rangle,
\end{equation}
that is, using $\langle z_j z_k\rangle= 0$ for $j\ne k$,
\begin{equation}
\langle P_j^2\rangle=
\langle 2\tilde{V}(z_j)+V(z_{j-1},z_j)+V(z_{j},z_{j+1})\rangle.
\end{equation}
We have checked numerically this latter equation.

In order to evaluate the equipartition times $\tau_p$ and $\tau_m$, we have
adopted the equipartition indicator given in Ref.~\cite{EPL99}. Namely, we
compute
$S_p (t)=\sum_{j=1}^{N_p} {\overline E}_{p,j}(t)$ $\ln {\overline E}_{p,j}(t)$
and
$S_m (t)=\sum_{k=1}^{N_m} {\overline E}_{m,k}(t) \ln {\overline E}_{m,k}(t)$,
where ${\overline E}_{p,j}(t)$ and ${\overline E}_{m,k}(t)$ are the time
averages of $P_j^2$ and $p_k^2$ up to time $t$, respectively. Then the value
$\exp[-S_p(t)]$ is a measure of the number of plates significantly excited up
to time $t$. We thus define $\tau_p$ as the time at which this value reaches
90\% of $N_p$. The same procedure is used to define $\tau_m$.

To measure numerically the pressure at a given time, a perturbation of
$-2\Delta l$ to the distance between the two mirrors is imposed; the
pressure is identified to be the ratio of the responding $\Delta E_{tot}$
over $2\Delta l$.

The fortran codes involved in this study are available. Those who are
interested, please contact the authors.


\end{document}